\documentclass[american]{article}
\usepackage[T1]{fontenc}
\usepackage[latin1]{inputenc}
\usepackage{geometry}
\usepackage{subfigure}
\usepackage{amsmath}
\usepackage{graphicx}
\usepackage{amssymb}

\makeatletter

\providecommand{\LyX}{L\kern-.1667em\lower.25em\hbox{Y}\kern-.125emX\@}

 \usepackage{verbatim}
 \newenvironment{lyxcode}
   {\begin{list}{}{
     \setlength{\rightmargin}{\leftmargin}
     \raggedright
     \setlength{\itemsep}{0pt}
     \setlength{\parsep}{0pt}
     \normalfont\ttfamily}%
    \item[]}
   {\end{list}}

\usepackage{fancyhdr}

\usepackage[nice]{nicefrac} 

\usepackage{babel}
\makeatother
\begin{document}

\title{Three-site quantum lattice with thermal bath }

\author{Paolo M. Pumilia}

\maketitle
\begin{center}paolo.pumilia@acm.org\end{center}

\date{august 19th, 2002}

\begin{abstract}
An algorithm to simulate the dynamics of a quantum state over a three-site
lattice interacting with classical harmonic oscillators has been devised. 

The oscillators are linearly coupled to the quantum state and are
acted upon by a fluctuation-dissipation process to take the equilibrium
thermal environment into account, thus allowing to investigate how
stochastic forces may affect the quantum dynamics.

The implementation of the algorithm has been written in \textsf{Ada95}.
\end{abstract}

\section{The physical model}

The Hosltein hamiltonian for a lattice connected to independent site
oscillators can be distinguished into three terms \begin{eqnarray}
h & = & h_{a}+h_{b}+h_{ab}\label{h}
\end{eqnarray}
corresponding to the \emph{primary} system $a$, representing those
states allowed to the quantum particle, the \emph{secondary} system
$b$, representing the vibrating lattice masses ($m$) and the interaction
energy $ab$ \cite{Holstein-59a}. 

In a three-site lattice, assuming the nearest neighbours approximation
hold, the quantum states are described by the hamiltonian \begin{equation}
h_{a}=V\left(\left|1\right\rangle \left\langle 2\right|+\left|2\right\rangle \left\langle 1\right|+\left|2\right\rangle \left\langle 3\right|+\left|3\right\rangle \left\langle 2\right|\right)\label{h_a}\end{equation}
 where $\left|n\right\rangle $ are the eigenstates of the isolated
sites and $V$ is the overlap integral between neighbour sites. 

Vibrations are described by harmonic oscillators of equal mass $m$
and frequency $\omega $, with respective position $r_{n}$ and conjugated
variable $p_{n}$, (where $n$ indicates the location on the lattice):\begin{equation}
h_{b}=\frac{1}{2m}\left(p_{1}^{2}+m^{2}\omega ^{2}r_{1}^{2}\right)+\frac{1}{2m}\left(p_{2}^{2}+m^{2}\omega ^{2}r_{2}^{2}\right)+\frac{1}{2m}\left(p_{3}^{2}+m^{2}\omega ^{2}r_{3}^{2}\right)\label{h_b}\end{equation}
Interaction terms are assumed to be linear with respect to $r_{n}$\begin{eqnarray}
h_{ab} & = & \epsilon _{1}\left|1\right\rangle \left\langle 1\right|+\epsilon _{2}\left|2\right\rangle \left\langle 2\right|+\epsilon _{3}\left|3\right\rangle \left\langle 3\right|\label{h_ab}
\end{eqnarray}
$\epsilon _{n}=\chi r_{n}$ , with $\chi $ real positive, being the
coupling energy at site $n$. Such physical model has been studued
by Hennig \cite{hennig-91} and Kenkre \& Andersen \cite{kenkre-dimer-86},
but without thermal environment.

\subsection{Lattice with free boundary conditions}

When loose ends boundary conditions apply to the primary system, we
are lead to the following equation for the density matrix: \begin{eqnarray}
\dot{\rho }_{11} & = & i\omega _{o}\left(\rho _{21}-\rho _{12}\right)\nonumber \\
\dot{\rho }_{12} & = & -i\omega _{12}\rho _{12}+i\omega _{o}\left(\rho _{22}-\rho _{11}-\rho _{13}\right)\nonumber \\
\dot{\rho }_{13} & = & -i\omega _{13}\rho _{13}+i\omega _{o}\left(\rho _{23}-\rho _{12}\right)\nonumber \\
\dot{\rho }_{21} & = & i\omega _{12}\rho _{21}+i\omega _{o}\left(\rho _{11}+\rho _{31}-\rho _{22}\right)\nonumber \\
\dot{\rho }_{22} & = & i\omega _{o}\left(\rho _{12}+\rho _{32}-\rho _{21}-\rho _{23}\right)\label{density-matrix-free}\\
\dot{\rho }_{23} & = & -i\omega _{23}\rho _{23}+i\omega _{o}\left(\rho _{13}+\rho _{33}-\rho _{22}\right)\nonumber \\
\dot{\rho }_{31} & = & i\omega _{13}\rho _{31}+i\omega _{o}\left(\rho _{21}-\rho _{32}\right)\nonumber \\
\dot{\rho }_{32} & = & i\omega _{23}\rho _{32}+i\omega _{o}\left(\rho _{22}-\rho _{31}-\rho _{33}\right)\nonumber \\
\dot{\rho }_{33} & = & i\omega _{o}\left(\rho _{23}-\rho _{32}\right)\nonumber 
\end{eqnarray}
where $\omega _{o}=V/h$ is the frequency for the oscillation of the
quasi-particle between two neighbour sites and $\omega _{kn}=-\omega _{nk}=\left(\epsilon _{n}-\epsilon _{k}\right)/\hbar =\chi \left(r_{n}-r_{k}\right)/\hbar $. 

Using operators \begin{eqnarray}
u_{1} & = & \rho _{11}-\rho _{22}\nonumber \\
u_{2} & = & \rho _{22}-\rho _{33}\nonumber \\
v_{1} & = & i\left(\rho _{12}-\rho _{21}\right)\nonumber \\
v_{2} & = & i\left(\rho _{32}-\rho _{23}\right)\label{density-operators}\\
v_{3} & = & i\left(\rho _{31}-\rho _{13}\right)\nonumber \\
w_{1} & = & \rho _{12}+\rho _{21}\nonumber \\
w_{2} & = & \rho _{23}+\rho _{32}\nonumber \\
w_{3} & = & \rho _{13}+\rho _{31}\nonumber 
\end{eqnarray}
a new set of equations, that will be used to perform numerical simulations,
can be obtained for the primary system. 

As a fluctuation-dissipation process has been finally attached to
each oscillator, the complete dynamical model can be written:\begin{eqnarray}
\dot{u}_{1} & = & -\omega _{o}\left(2v_{1}+v_{2}\right)\nonumber \\
\dot{u}_{2} & = & \omega _{o}\left(v_{1}+2v_{2}\right)\nonumber \\
\dot{v}_{1} & = & \omega _{12}w_{1}+\omega _{o}\left(2u_{1}+w_{3}\right)\nonumber \\
\dot{v}_{2} & = & -\omega _{23}w_{2}-\omega _{o}\left(2u_{2}-w_{3}\right)\nonumber \\
\dot{v}_{3} & = & -\omega _{13}w_{3}-\omega _{o}\left(w_{1}-w_{2}\right)\nonumber \\
\dot{w}_{1} & = & -\omega _{12}v_{1}+\omega _{o}v_{3}\nonumber \\
\dot{w}_{2} & = & \omega _{23}v_{2}-\omega _{o}v_{3}\nonumber \\
\dot{w}_{3} & = & \omega _{13}v_{3}-\omega _{o}\left(v_{1}+v_{2}\right)\label{sch eq with noise}\\
\dot{r}_{1} & = & p_{1}/m\nonumber \\
\dot{r}_{2} & = & p_{2}/m\nonumber \\
\dot{r}_{3} & = & p_{3}/m\nonumber \\
\dot{p}_{1} & = & -m\omega ^{2}r_{1}-\frac{\chi }{3}\left(c+u_{2}+2u_{1}\right)-\gamma _{1}p_{1}+f_{1}\left(t\right)\nonumber \\
\dot{p}_{2} & = & -m\omega ^{2}r_{2}-\frac{\chi }{3}\left(c+u_{2}-u_{1}\right)-\gamma _{2}p_{2}+f_{2}\left(t\right)\nonumber \\
\dot{p}_{3} & = & -m\omega ^{2}r_{3}-\frac{\chi }{3}\left(c-u_{1}-2u_{2}\right)-\gamma _{3}p_{3}+f_{3}\left(t\right)\nonumber 
\end{eqnarray}
where $\gamma _{n}$ are the damping coefficients and $f_{n}\left(t\right)$
models gaussian noise, with $\delta $-shaped time correlation, satisfying
the following relations:\begin{eqnarray}
\left\langle f_{n}(t)\right\rangle  & = & 0\label{zero mean f}\\
\left\langle f_{n}(t)f_{n}(t')\right\rangle  & = & 2\gamma \theta \delta _{\left(t-t'\right)}\label{delta corr f}
\end{eqnarray}
having defined thermal energy\begin{equation}
\theta =k_{B}T\label{thermal energy}\end{equation}
($k_{B}$being the Boltzmann constant); hence \begin{equation}
\left\langle f_{n}^{2}(t)\right\rangle ^{1/2}=\left(2\gamma \theta \right)^{1/2}\label{variance f}\end{equation}

It can be readily verified that \begin{equation}
c=\rho _{11}+\rho _{22}+\rho _{33}\label{const 1}\end{equation}
is a constant of motion. A further constant, that will be used to
monitor numerical simulations, is given by \begin{equation}
K=\frac{4}{3}\left(u_{1}^{2}+u_{2}^{2}+u_{1}u_{2}\right)+v_{1}^{2}+v_{2}^{2}+v_{3}^{2}+w_{1}^{2}+w_{2}^{2}+w_{3}^{2}\label{const 2}\end{equation}

The dynamical problem is described by a system of stochastic real
valued, first order, differential equations (\ref{sch eq with noise})
that must be numerically solved.

Notice that, restricting the number of sites to $n=1,2$ we are left
with the spin-boson equation with noise: \begin{eqnarray}
\dot{u}_{1} & = & -2\omega _{o}v_{1}\nonumber \\
\dot{v}_{1} & = & \omega _{12}w_{1}+2\omega _{o}u_{1}\nonumber \\
\dot{w}_{1} & = & -\omega _{12}v_{1}\nonumber \\
\dot{r}_{1} & = & p_{1}/m\label{spin-boson with noise}\\
\dot{r}_{2} & = & p_{2}/m\nonumber \\
\dot{p}_{1} & = & -m\omega ^{2}r_{1}-\frac{\chi }{3}\left(c+u_{2}+2u_{1}\right)+f_{1}\left(t\right)\nonumber \\
\dot{p}_{2} & = & -m\omega ^{2}r_{2}-\frac{\chi }{3}\left(c+u_{2}-u_{1}\right)+f_{2}\left(t\right)\nonumber 
\end{eqnarray}
that has been studied by many researchers; in particular, the author's
work has been inspired the papers\cite{kenkre-dimer-86,grigo-dimer-91},
referenced to in the bibliography.

\section{Numerical integration method}

Stochastic differential equations of first order in time for a vector
variabile $\mathbf{x}(t)$ take the from, \begin{equation}
\dot{\mathbf{x}}=\mathbf{F}+\mathbf{gf}\label{eq_diff}\end{equation}
where $\mathbf{F=}\left(F_{i}\right)_{1,n}=\left(F_{i}(\mathbf{x},t)\right)_{1,n}$
is the vector of deterministic fields, $\mathbf{f=}\left(f_{i}\right)_{1,n}=\left(f_{i}(t)\right)_{1,n}$
the gaussian stochastic force, satisfying the conditions:\begin{eqnarray}
\left\langle f(t)\right\rangle  & = & 0\nonumber \\
\left\langle f(t)f(t')\right\rangle  & = & \delta (t-t')\label{gaussian_noise}
\end{eqnarray}
while th parameters set $\mathbf{g=}\left(g_{i}\right)_{1,n}=\left(g_{i}\left(\mathbf{x},t\right)\right)_{1,n}$
represents the interaction of the system with the thrmal bath. 

Equations given in (\ref{eq_diff}) are \emph{entirely coupled,} since,
in general, every components $F_{i}$ and $g_{i}$ are functions of
the whole variable $\mathbf{x}$,

The formal solution of (\ref{eq_diff}), assuming the \emph{implicit}
time dependence $F=\mathbf{F}(\mathbf{x}(t))$ e $g=\mathbf{g}(\mathbf{x}(t))$,
is given by \cite{spde}\begin{eqnarray}
x_{i}(t)-x_{i}(0)=\int _{0}^{t}dt'F_{i}+\int _{0}^{t}dt'g_{i}f_{i} & \, \, \, \, \, \, \, \, \,  & i=1,\ldots n\label{formal_int}
\end{eqnarray}

Given a time interval $\left[0,h\right]$ that can be considered infinitely
small for $F_{i}$ e $g_{i}$ functions, their value at the arbitrary
instant $t'\in \left[0,h\right]$, can be approximated by the first
$\kappa $ terms of a Taylor expansion, centered at $t=0$:\begin{eqnarray}
F_{i}\left(t\right) & = & F_{i}^{o}+\delta ^{\kappa }F_{i}\label{F series}\\
g_{i}\left(t\right) & = & g_{i}^{o}+\delta ^{\kappa }g_{i}\label{g series}
\end{eqnarray}
The index $\kappa $ in (\ref{F series},\ref{g series}) indicates
the truncation order of the series: \begin{equation}
\delta F_{i}=\sum _{j}\left[F_{i}\right]_{j}\cdot \delta x_{j}(t')+\frac{1}{2}\sum _{jk}\left[F_{i}\right]_{jk}\cdot \delta x_{j}(t')\delta x_{k}(t')+\frac{1}{3!}\sum _{jk}\left[F_{i}\right]_{jkl}\cdot \delta x_{j}(t')\delta x_{k}(t')\delta x_{l}+\ldots \label{dF_series}\end{equation}
\begin{equation}
\delta g_{i}=\sum _{j}\left[g_{i}\right]_{j}\cdot \delta x_{j}(t')+\frac{1}{2}\sum _{jk}\left[g_{i}\right]_{jk}\cdot \delta x_{j}(t')\delta x_{k}(t')+\frac{1}{3!}\sum _{jk}\left[g_{i}\right]_{jkl}\cdot \delta x_{j}(t')\delta x_{k}(t')\delta x_{l}+\ldots \label{dg_series}\end{equation}
in which $F_{i}$ and $g_{i}$, time derivatives evaluated at $t=0$,
have been represented by the following notation:\begin{equation}
\left[F_{i}\right]_{j}\equiv \frac{\partial F_{i}}{\partial x_{j}}(0)\, \, \, \, \, \, \left[F_{i}\right]_{jk}\equiv \frac{\partial ^{2}F_{i}}{\partial x_{j}\partial x_{k}}(0)\, \, \, \, \, \, \left[F_{i}\right]_{jkl}\equiv \frac{\partial ^{3}F_{i}}{\partial x_{j}\partial x_{k}\partial x_{l}}(0)\label{dF_coeff}\end{equation}
\begin{equation}
\left[g_{i}\right]_{j}\equiv \frac{\partial g_{i}}{\partial x_{j}}(0)\, \, \, \, \, \, \left[g_{i}\right]_{jk}\equiv \frac{\partial ^{2}g_{i}}{\partial x_{j}\partial x_{k}}(0)\, \, \, \, \, \, \left[g_{i}\right]_{jkl}\equiv \frac{\partial ^{3}g_{i}}{\partial x_{j}\partial x_{k}\partial x_{l}}(0)\label{dg_coeff}\end{equation}

Truncation at $\kappa =0$ the general solution obtained after substitution
of (\ref{F series},\ref{g series}) into (\ref{formal_int}):\begin{equation}
x_{i}(h)-x_{i}(0)=\int _{0}^{h}dt'\left(F_{i}^{o}+\delta ^{\kappa }F_{i}\right)+\int _{0}^{h}dt'\left(g_{i}^{o}+\delta ^{\kappa }g_{i}\right)f_{i}\label{formal_int_series}\end{equation}
 yields\begin{equation}
x_{i}(h)-x_{i}(0)=F_{i}^{o}h+g_{i}^{o}\int _{0}^{h}dt'f_{i}\label{zero_int}\end{equation}
where $F_{i}=\left[F_{i}\right]=F_{i}^{o}$ and $g_{i}=\left[g_{i}\right]=g_{i}^{o}$
have been defined.

Since the gaussian integral \begin{equation}
Z_{1}(h)=\int _{0}^{h}dt'f(t')\label{Z1}\end{equation}
 is of $h^{1/2}$ order, the second term in (\ref{zero_int}) is the
lowest order approximation of the trajectory (\ref{formal_int_series}):\begin{equation}
\delta x_{i}^{(1/2)}=g_{i}^{o}Z_{1}\label{order 1/2}\end{equation}
 having defined $\delta x_{i}=x_{i}(h)-x_{i}(0)$. 

Substituting (\ref{order 1/2}) into (\ref{dF_series},\ref{dg_series}),
hence into (\ref{formal_int}), and retaining only contribution of
the order $h$ in (\ref{formal_int_series}), we obtain the first
order correction

\begin{eqnarray}
\delta x_{i}^{(1)} & = & F_{i}^{o}h+\int _{0}^{h}dt'\sum _{j}\left[g_{i}\right]_{j}\cdot \delta x_{j}^{(1/2)}f_{i}(t')\nonumber \\
 & = & F_{i}^{o}h+\sum _{j}\left[g_{i}\right]_{j}\cdot g_{j}^{o}\int _{0}^{h}dt'Z_{1}(t')f_{i}(t')\nonumber \\
 & = & F_{i}^{o}h+\frac{1}{2}\sum _{j}\left[g_{i}\right]_{j}\cdot g_{j}^{o}Z_{1}^{2}\label{order 1 correction}
\end{eqnarray}

The corresponding displacement on the trajectory is thus given by\begin{equation}
\delta x_{i}=\delta x_{i}^{1/2}+\delta x_{i}^{1}\label{order 1}\end{equation}
Again substituting the first order correction (\ref{order 1 correction})
into the series (\ref{dF_series},\ref{dg_series}), hence into (\ref{formal_int}),
then retaining only terms of the order $h^{3/2}$ in (\ref{formal_int_series}),
we get:\begin{eqnarray}
\delta x_{i}^{(3/2)} & = & \sum _{j}\left[F_{i}\right]_{j}\cdot g_{j}^{o}Z_{1}h+\int _{0}^{h}dt'\sum _{j}\left[g_{i}\right]_{j}\cdot \delta x_{j}^{(1)}f_{i}(t')\nonumber \\
 & = & \sum _{j}\left[F_{i}\right]_{j}\cdot g_{j}^{o}Z_{1}h+\int _{0}^{h}dt'\sum _{j}\left[g_{i}\right]_{j}\cdot \left(F_{i}^{o}h+\frac{1}{2}\sum _{k}\left[g_{i}\right]_{k}\cdot g_{k}^{o}Z_{1}^{2}(t')f_{i}(t')\right)\nonumber \\
 & = & \sum _{j}\left[F_{i}\right]_{j}\cdot g_{j}^{o}Z_{1}h+\sum _{j}\left[g_{i}\right]_{j}\cdot \left(F_{i}^{o}h+\frac{1}{2}\sum _{k}\left[g_{i}\right]_{k}\cdot g_{k'}^{o}\int _{0}^{h}dtZ_{1}^{2}(t')f_{i}(t')\right)\nonumber \\
 & = & \sum _{j}\left[F_{i}\right]_{j}\cdot g_{j}^{o}Z_{1}h+\sum _{j}\left[g_{i}\right]_{j}\cdot \left(F_{i}^{o}h+\frac{1}{3!}\sum _{k}\left[g_{i}\right]_{k}\cdot g_{k'}^{o}Z_{1}^{3}\right)\label{order 3/2}
\end{eqnarray}

In the simplified case in which $g$ does not depend on $\mathbf{x}$,
the numerical algorithm to reach the order $h^{2}$ in approximating
the exact solution of (\ref{eq_diff}) can then be written\begin{equation}
x_{i}(h)-x_{i}(0)=g_{i}^{o}Z_{1}+F_{i}^{o}h+\sum _{j}\left[F_{i}\right]_{j}\cdot g_{j}^{o}Z_{2}+\frac{1}{2}\sum _{j}\left[F_{i}\right]_{j}\cdot \left[F_{j}\right]h^{2}\, \, \, \, \, \, \, \, \, i=1,\ldots n\label{order 2 solution}\end{equation}
or, in vectorial notation:\begin{equation}
\mathbf{x}(h)-\mathbf{x}(0)=\mathbf{g}^{o}\mathbf{Z}_{1}+\mathbf{F}^{o}\mathbf{h}+\nabla \mathbf{FgZ}_{2}+\frac{1}{2}\nabla \mathbf{F}\cdot \mathbf{F}h^{2}\label{order 2 vect sol}\end{equation}

\subsection{The algorithm for the three-site lattice}

Equations (\ref{sch eq with noise}) belong to the class (\ref{eq_diff});
therefore the approximation method that has been recalled in the previous
chapter can be implemented. 

To better expose our procedure, it is convenient switching to the
vectorial notation; the system variable $\mathbf{x}$, taking the
form:\begin{equation}
\mathbf{x}=\left[\begin{array}{c}
 u_{1}\\
 u_{2}\\
 v_{1}\\
 v_{2}\\
 v_{3}\\
 w_{1}\\
 w_{2}\\
 w_{3}\\
 r_{1}\\
 r_{2}\\
 r_{3}\\
 p_{1}\\
 p_{2}\\
 p_{3}\end{array}
\right]\label{variables}\end{equation}
 evolves in time according to the equation\begin{equation}
\dot{\mathbf{x}}=\mathbf{F}+\mathbf{G}\label{vectorial dyn eq.}\end{equation}
where $\mathbf{F}$ is the vector of the deterministic forces \begin{eqnarray}
\mathbf{F} & = & \left[\begin{array}{c}
 -\omega _{o}\left(2v_{1}+v_{2}\right)\\
 \omega _{o}\left(v_{1}+2v_{2}\right)\\
 \omega _{12}w_{1}+\omega _{o}\left(2u_{1}+w_{3}\right)\\
 -\omega _{23}w_{2}-\omega _{o}\left(2u_{2}-w_{3}\right)\\
 -\omega _{13}w_{3}-\omega _{o}\left(w_{1}-w_{2}\right)\\
 -\omega _{12}v_{1}+\omega _{o}v_{3}\\
 \omega _{23}v_{2}-\omega _{o}v_{3}\\
 \omega _{13}v_{3}-\omega _{o}\left(v_{1}+v_{2}\right)\\
 p_{1}/m\\
 p_{2}/m\\
 p_{3}/m\\
 -m\omega ^{2}r_{1}-\frac{\chi }{3}\left(c+u_{2}+2u_{1}\right)-\gamma _{1}p_{1}\\
 -m\omega ^{2}r_{2}-\frac{\chi }{3}\left(c+u_{2}-u_{1}\right)-\gamma _{2}p_{2}\\
 -m\omega ^{2}r_{3}-\frac{\chi }{3}\left(c-u_{1}-2u_{2}\right)-\gamma _{3}p_{3}\end{array}
\right]\label{deterministic forces}
\end{eqnarray}
while \textbf{$\mathbf{G}$} contains the average values of the stochastic
forces:\begin{equation}
\mathbf{G}=\left[\begin{array}{c}
 0\\
 0\\
 0\\
 0\\
 0\\
 0\\
 0\\
 0\\
 0\\
 0\\
 0\\
 \left(2\gamma _{1}\theta _{1}\right)^{1/2}\\
 \left(2\gamma _{2}\theta _{2}\right)^{1/2}\\
 \left(2\gamma _{3}\theta _{3}\right)^{1/2}\end{array}
\right]\label{averaged stochastic forces}\end{equation}

The approximated solution up to the second order in time can be computed
according to the formula (\ref{order 2 vect sol}):\begin{equation}
\mathbf{x}(h)\simeq \mathbf{x}(0)+\mathbf{g}\cdot \mathbf{Z}_{1}+\mathbf{F}\cdot \mathbf{h}+\nabla \mathbf{F}\cdot \mathbf{g}\cdot \mathbf{Z}_{2}+\frac{1}{2}\nabla \mathbf{F}\cdot \mathbf{F}\cdot \mathbf{h}^{2}\label{order 2 integration}\end{equation}
 where $Z_{1}$ and $Z_{2}$ are gaussian integrals and the force
gradient is given by:\begin{equation}
\nabla \mathbf{F}=\left[\begin{array}{cccccccccccccc}
 0 & 0 & -2\omega _{o} & -\omega _{o} & 0 & 0 & 0 & 0 & 0 & 0 & 0 & 0 & 0 & 0\\
 0 & 0 & \omega _{o} & 2\omega _{o} & 0 & 0 & 0 & 0 & 0 & 0 & 0 & 0 & 0 & 0\\
 2\omega _{o} & 0 & 0 & 0 & 0 & \omega _{12} & 0 & \omega _{o} & -\chi _{1}w_{1} & \chi _{2}w_{1} & 0 & 0 & 0 & 0\\
 0 & -2\omega _{o} & 0 & 0 & 0 & 0 & -\omega _{23} & \omega _{o} & 0 & \chi _{2}w_{2} & -\chi _{3}w_{2} & 0 & 0 & 0\\
 0 & 0 & 0 & 0 & 0 & -\omega _{o} & \omega _{o} & -\omega _{13} & \chi _{1}w_{3} & 0 & -\chi _{3}w_{3} & 0 & 0 & 0\\
 0 & 0 & -\omega _{12} & 0 & \omega _{o} & 0 & 0 & 0 & \chi _{1}v_{1} & -\chi _{2}v_{1} & 0 & 0 & 0 & 0\\
 0 & 0 & 0 & \omega _{23} & -\omega _{o} & 0 & 0 & 0 & 0 & -\chi _{1}v_{2} & \chi _{3}v_{2} & 0 & 0 & 0\\
 0 & 0 & -\omega _{o} & -\omega _{o} & \omega _{13} & 0 & 0 & 0 & -\chi _{1}v_{3} & 0 & \chi _{3}v_{3} & 0 & 0 & 0\\
 0 & 0 & 0 & 0 & 0 & 0 & 0 & 0 & 0 & 0 & 0 & \nicefrac 1m_{1} & 0 & 0\\
 0 & 0 & 0 & 0 & 0 & 0 & 0 & 0 & 0 & 0 & 0 & 0 & \nicefrac 1m_{2} & 0\\
 0 & 0 & 0 & 0 & 0 & 0 & 0 & 0 & 0 & 0 & 0 & 0 & 0 & \nicefrac 1m_{2}\\
 -2\chi _{1}/3 & -\chi _{1}/3 & 0 & 0 & 0 & 0 & 0 & 0 & -m_{1}\omega _{1}^{2} & 0 & 0 & -\gamma _{1} & 0 & 0\\
 \chi _{2}/3 & -\chi _{2}/3 & 0 & 0 & 0 & 0 & 0 & 0 & 0 & -m_{2}\omega _{2}^{2} & 0 & 0 & -\gamma _{2} & 0\\
 \chi _{3}/3 & 2\chi _{3}/3 & 0 & 0 & 0 & 0 & 0 & 0 & 0 & 0 & -m_{3}\omega _{3}^{2} & 0 & 0 & -\gamma _{3}\end{array}
\right]\end{equation}

The implementation of the algorithm according to the rule (\ref{order 2 integration})
is displayed in the last section . \textsf{Ada95} language has been
the language of choiche, since it is maintained that a rather easy
and dependable parallel code, that could be required to perform long
lasting simulations can be worked out in that environment, .

Marsaglia \& Tsang \cite{marsaglia} fortran routine for random numbers
generation has been used to compute $Z_{1}$ and $Z_{2}$ integrals.

\section{Results and Conclusions}

First run of the program allowed to plot variables {\scriptsize :}
$u_{1}=\rho _{11}-\rho _{22}$ and $u_{2}=\rho _{22}-\rho _{33}$,
using the set of physical parameters:

\begin{itemize}
\item inter-site interaction $V=0.09$ 
\item free oscillators frequencies $\omega ^{2}=2x10^{-3}$ 
\item vibration-quantum state coupling $\chi =4x10^{-2}$ ; 
\item $\gamma =0.2$ that imply $\omega ^{2}/2\gamma ^{2}=0.025$ and $\chi ^{2}/2m\omega ^{2}=0.1$ 
\end{itemize}
Time integration step has been set to $10^{-3}$ and the reliability
of the computation has been monitored through the constant (\ref{const 2}),
whose changes have kept below $10^{-7}$.

\begin{figure}[hbtp]
\subfigure[occupation differences  $u_{1}=\rho _{11}-\rho _{22}$ (lower) and $u_{2}=\rho _{22}-\rho _{33}$ (upper) evolving in time]{\includegraphics[  width=16cm,
  height=9cm]{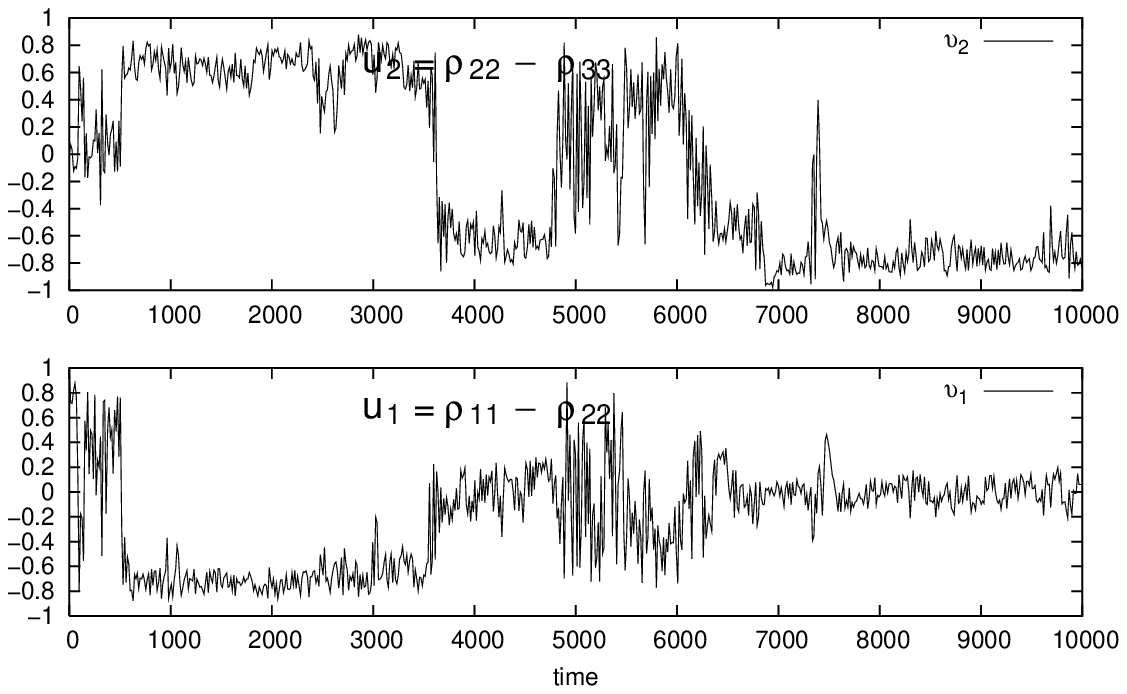}}
\end{figure}

Picture of the site occupation differences $u_{1}=\rho _{11}-\rho _{22}$
(lower) and $u_{2}=\rho _{22}-\rho _{33}$ (upper) evolving in time
(arbitrary units) shows the localized motion of the quantum state.

The possibilty to model the dynamics of a three-site lattice coupled
to stochastic oscillators allows one to study how a quantum state
propagates in a real lattice, when vibrational motion can be described
classically, then to investigate the duration of out of equilibrium
states and the conditions under which energy transfer without dissipation
may occur.

\section{Code listing}

Here included are the main tokens of code from the \textsf{\small iterator}
procedure, in which the numerical integration at time step \texttt{\small dh}
is performed:

\begin{lyxcode}
{\footnotesize procedure~iterator~(x~:~in~out~coordinates~;~xh~:~out~coordinates~;~~~~~~~~~~~~~~~~~~~~~~~~~~~~~~~~~~~~~~~~~~~~~time\_step~~~:~in~double~;~~-{}-~time~integration~step~~~~~~~~~~~~~~~~~~~~~~~~~~~~~~~~~~~~~~~~~~inner\_cycle~:~in~integer~;~-{}-~number~of~steps~between~next~-{}-~print~outs}{\footnotesize \par}

~{\footnotesize ~~~~~~~~~~~~~~~~~~~outer\_cycle~:~in~integer~;~-{}-~total~number~of~trajectory~~~-{}-~points~to~print~out~}{\footnotesize \par}

~{\footnotesize ~~~~~~~~~~~~~~~~~~~v,am1,am2,am3,gam1,gam2,gam3,om1,om2,om3:~in~double;~~~~~~~~~~~~~~~~~~~~~~~~~~~~~~~~~~~~~~~~chi1,chi2,chi3,t1,t2,t3,e1b,e2b,e3b~~:~in~double;~~~~~~~~~~~~~~}{\footnotesize \par}

~{\footnotesize ~~~~~~~~~~~~~~~~~~~output\_file~:~~in~string~~~~~~~~~~~~~~~~)~}{\footnotesize \par}

{\footnotesize is~~}{\footnotesize \par}

\bigskip{}
\begin{flushleft}\textrm{Fortran random numbers generation routine}
\textsf{\small rnor} \textrm{is imported here:}\end{flushleft}
\medskip{}

{\footnotesize function~rnor(dum~:~in~Fortran.Fortran\_Integer)~return~Fortran.double\_precision;~~}{\footnotesize \par}

{\footnotesize pragma~Import~(fortran,~rnor,~\char`\"{}rnor\_\char`\"{});~}{\footnotesize \par}

\bigskip{}
\begin{flushleft}\textrm{Definition of numerical parameters:}\end{flushleft}
\medskip{}

{\footnotesize dh~~:~double~:=~time\_step~;~-{}-~time~integratin~step~~~~}{\footnotesize \par}

{\footnotesize dh2~:~double~:=~time\_step~{*}~time\_step;~-{}-~~squared~time~integration~step~}{\footnotesize \par}

{\footnotesize dh25~:=~dh2~{*}~0.5;~~}{\footnotesize \par}

\bigskip{}
\textrm{Definition~of~physical~parameters:}
\medskip{}

{\footnotesize v2:~~~double~:=~v~{*}~v~;~-{}-~~}{\footnotesize \par}

{\footnotesize ak0~:=~c1{*}c1~+~c2{*}c2~+~c3{*}c3;~~~~~~~~}{\footnotesize \par}

{\footnotesize qom1~:=~om1{*}om1;~}{\footnotesize \par}

{\footnotesize qom2~:=~om2{*}om2;~~}{\footnotesize \par}

{\footnotesize qom3~:=~om3{*}om3;}{\footnotesize \par}

{\footnotesize qgam1~:=~gam1~{*}~gam1;~~}{\footnotesize \par}

{\footnotesize qgam2~:=~gam2~{*}~gam2;~}{\footnotesize \par}

{\footnotesize qgam3~:=~gam3~{*}~gam3;}{\footnotesize \par}

{\footnotesize norm~:=~(1.0/3.0){*}(~x(1){*}x(1)~+~x(2){*}x(2)~+~x(1){*}x(2)~);~}{\footnotesize \par}

{\footnotesize norm~:=~norm~+~x(3){*}x(3)~+~x(4){*}x(4)~+~x(5){*}x(5);}{\footnotesize \par}

{\footnotesize norm~:=~norm~+~x(6){*}x(6)~+~x(7){*}x(7)~+~x(8){*}x(8);~}{\footnotesize \par}

{\footnotesize norm~:=~(3.0/4.0)~{*}~norm;}{\footnotesize \par}

\bigskip{}
\textrm{Here  the iteration starts:}
\medskip{}

{\footnotesize for~i~in~1..outer\_cycle~loop~~~~~~~~~~~~~~}{\footnotesize \par}

~{\footnotesize ~for~j~in~1..inner\_cycle~loop~~~~~~}{\footnotesize \par}

\bigskip{}
\textrm{Definition~of~some~useful~expressionss:}
\medskip{}

~{\footnotesize ~~~time~:=~time~+~time\_step;}{\footnotesize \par}

~{\footnotesize ~~~w11~:=~double(rnor(dum));~~~~}{\footnotesize \par}

~{\footnotesize ~~~w12~:=~double(rnor(dum));~~~~~~}{\footnotesize \par}

~{\footnotesize ~~~w21~:=~double(rnor(dum));~~~~~~~}{\footnotesize \par}

~{\footnotesize ~~~w22~:=~double(rnor(dum));~~~~~~}{\footnotesize \par}

~{\footnotesize ~~~w31~:=~double(rnor(dum));~~~~~}{\footnotesize \par}

~{\footnotesize ~~~w32~:=~double(rnor(dum));~~~~~}{\footnotesize \par}

~{\footnotesize ~~~z11~:=~a11{*}w11~;~}{\footnotesize \par}

~{\footnotesize ~~~z12~:=~a21{*}z11~+~a22{*}w12;~~}{\footnotesize \par}

~{\footnotesize ~~~z21~:=~a11{*}w21~;~~~~}{\footnotesize \par}

~{\footnotesize ~~~z22~:=~a21{*}z21~+~a22{*}w22;~~~~}{\footnotesize \par}

~{\footnotesize ~~~z31~:=~a11{*}w31~;~~~~~}{\footnotesize \par}

~{\footnotesize ~~~z32~:=~a21{*}z31~+~a22{*}w32;}{\footnotesize \par}

~{\footnotesize ~~~om12~:=~de12~~-~chi2{*}x(10)~+~chi1{*}x(9);~~~}{\footnotesize \par}

~{\footnotesize ~~~om23~:=~de23~~-~chi3{*}x(11)~+~chi2{*}x(10);~~~~~~~~}{\footnotesize \par}

~{\footnotesize ~~~om13~:=~de13~~-~chi3{*}x(11)~+~chi1{*}x(9);~~~~~~~~~~~~}{\footnotesize \par}

~{\footnotesize ~~~qom12~:=~om12~{*}~om12;~~~~~~~~}{\footnotesize \par}

~{\footnotesize ~~~qom13~:=~om13~{*}~om13;~~~~~}{\footnotesize \par}

~{\footnotesize ~~~qom23~:=~om23~{*}~om23;~~~~~~~~}{\footnotesize \par}

~{\footnotesize ~~~akey21~:=~-~chi2{*}x(13)/am2~+~chi1{*}x(12)/am1~;~~~~~~}{\footnotesize \par}

~{\footnotesize ~~~akey23~:=~-~chi2{*}x(13)/am2~+~chi3{*}x(14)/am3~;~~~~~}{\footnotesize \par}

~{\footnotesize ~~~akey31~:=~-~chi3{*}x(14)/am3~+~chi1{*}x(12)/am1~;}{\footnotesize \par}

~{\footnotesize ~~~ajay1~:=~chi1{*}(ak0+x(1)/2.0+x(2))/3.0;~~~~~~~}{\footnotesize \par}

~{\footnotesize ~~~ajay2~:=~chi2{*}(ak0+x(1)/2.0-x(2)/2.0)/3.0;~~~~~~~}{\footnotesize \par}

~{\footnotesize ~~~ajay3~:=~chi3{*}(ak0-x(1)-x(2)/2.0)/3.0;~~~~~~~~~~}{\footnotesize \par}

~{\footnotesize ~~~omx3~:=~om23~+~om13;~~~~~~~~}{\footnotesize \par}

~{\footnotesize ~~~omx1~:=~om12~+~om13;~}{\footnotesize \par}

\bigskip{}
\textrm{Variables~incrementation:}
\medskip{}

~{\footnotesize ~~~xh(1)~:=~x(1)~+~2.0{*}v{*}(~2.0{*}x(7)~+~x(6)~)~{*}~dh~+~~}{\footnotesize \par}

~{\footnotesize ~~~~~~~~~~~~~~~~~~~v2{*}(x(2)-2.0{*}x(1)+3.0{*}x(5))~{*}~dh2~+~~~~~~~~~~}{\footnotesize \par}

~{\footnotesize ~~~~~~~~~~~~~~~~~~~v{*}(2.0{*}om23{*}x(4)~-~om12{*}x(3))~{*}~dh2;}{\footnotesize \par}

~

~{\footnotesize ~~~xh(2)~:=~x(2)~-~2.0{*}v{*}(~x(7)~+~2.0{*}x(6)~)~{*}~dh~+~}{\footnotesize \par}

~{\footnotesize ~~~~~~~~~~~~~~~~~~~v2{*}(x(1)-2.0{*}x(2)-3.0{*}x(5))~{*}~dh2~+~~~~~~~}{\footnotesize \par}

~{\footnotesize ~~~~~~~~~~~~~~~~~~~v{*}(2.0{*}om12{*}x(3)~-~om23{*}x(4))~{*}~dh2;}{\footnotesize \par}

~

~{\footnotesize ~~~xh(3)~:=~x(3)~+~(~om12{*}x(6)~+~v{*}x(8)~)~{*}~dh~-~~~}{\footnotesize \par}

~{\footnotesize ~~~~~~~~~~~~~~~~~~~~~qom12{*}x(3)~{*}~dh25~+~v2{*}(x(4)-x(3))~{*}~dh25~+~~~~~~~}{\footnotesize \par}

~{\footnotesize ~~~~~~~~~~~~~~~~~~~~~v{*}(om12{*}x(2)~+~omx1{*}x(5))~{*}~dh25~+~~~~~~~~~~~~~~~}{\footnotesize \par}

~{\footnotesize ~~~~~~~~~~~~~~~~~~~~~akey21{*}x(6)~{*}~dh25~;}{\footnotesize \par}

~

~{\footnotesize ~~~xh(4)~:=~x(4)~-~(~om23{*}x(7)~+~v{*}x(8)~)~{*}~dh~-~~}{\footnotesize \par}

~{\footnotesize ~~~~~~~~~~~~~~~~~~~~~qom23{*}x(4)~{*}~dh25~+~v2{*}(x(3)-x(4))~{*}~dh25~+~~~~~~~~~~~~~~~~}{\footnotesize \par}

~{\footnotesize ~~~~~~~~~~~~~~~~~~~~~v{*}(om23{*}x(1)~-~omx3{*}x(5))~{*}~dh25~+~~}{\footnotesize \par}

~{\footnotesize ~~~~~~~~~~~~~~~~~~~~~akey23{*}x(7)~{*}~dh25~;}{\footnotesize \par}

~

~{\footnotesize ~~~xh(5)~:=~x(5)~-~(~om13{*}x(8)~+~v{*}x(6)~+~v{*}x(7)~)~{*}~dh~-~~~}{\footnotesize \par}

~{\footnotesize ~~~~~~~~~~~~~~~~~~~~~qom13{*}x(5)~{*}~dh25~+~v2{*}(x(1)-x(2)~-~2.0{*}x(5))~{*}~dh25~+~~~~~~~~~~~~~~~~}{\footnotesize \par}

~{\footnotesize ~~~~~~~~~~~~~~~~~~~~~v{*}(omx1{*}x(3)~-~omx3{*}x(4))~{*}~dh25~-~~~}{\footnotesize \par}

~{\footnotesize ~~~~~~~~~~~~~~~~~~~~~akey31{*}x(8)~{*}~dh25~;}{\footnotesize \par}

~

~{\footnotesize ~~~xh(6)~:=~x(6)~-~(~om12{*}x(3)~-~v{*}x(2)~-~v{*}x(5)~)~{*}~dh~-}{\footnotesize \par}

~{\footnotesize ~~~~~~~~~~~~~~~~~~~~~qom12{*}x(6)~{*}~dh25~-~v2{*}(5.0{*}x(6)+3.0{*}x(7))~{*}~dh25~-~~~~~~~~~~~~~~~~}{\footnotesize \par}

~{\footnotesize ~~~~~~~~~~~~~~~~~~~~~v{*}omx1{*}x(8)~{*}~dh25~-~~~}{\footnotesize \par}

~{\footnotesize ~~~~~~~~~~~~~~~~~~~~~akey21{*}x(3)~{*}~dh25~;}{\footnotesize \par}

~

~{\footnotesize ~~~xh(7)~:=~x(7)~+~(~om23{*}x(4)~-~v{*}x(1)~+~v{*}x(5)~)~{*}~dh~-~}{\footnotesize \par}

~{\footnotesize ~~~~~~~~~~~~~~~~~~~~~qom23{*}x(7)~{*}~dh25~-~v2{*}(3.0{*}x(6)+5.0{*}x(7))~{*}~dh25~-~~~~~~~~~~~~~~~~}{\footnotesize \par}

~{\footnotesize ~~~~~~~~~~~~~~~~~~~~~v{*}omx3{*}x(8)~{*}~dh25~-~~~}{\footnotesize \par}

~{\footnotesize ~~~~~~~~~~~~~~~~~~~~~akey23{*}x(4)~{*}~dh25~;}{\footnotesize \par}

~

~{\footnotesize ~~~xh(8)~:=~x(8)~+~(~om13{*}x(5)~-~v{*}x(3)~+~v{*}x(4)~)~{*}~dh~-~~~~~~~~~~~}{\footnotesize \par}

~{\footnotesize ~~~~~~~~~~~~~~~~~~~~~qom13{*}x(8)~{*}~dh25~-~2.0{*}v2{*}x(8)~{*}~dh25~-~~}{\footnotesize \par}

~{\footnotesize ~~~~~~~~~~~~~~~~~~~~~v{*}(omx1{*}x(6)~+~omx3{*}x(7))~{*}~dh25~+~~~~~~~~~~~~~~~~}{\footnotesize \par}

~{\footnotesize ~~~~~~~~~~~~~~~~~~~~~akey31{*}x(5)~{*}~dh25~;}{\footnotesize \par}

~

~{\footnotesize ~~~xh(9)~:=~x(9)~+~(x(12)/am1)~{*}~dh~-~~~~~~~~~~~~~~~~~}{\footnotesize \par}

~{\footnotesize ~~~~~~~~~~~~~~~~~~~~qom1{*}x(9)~{*}~dh25~-~~~}{\footnotesize \par}

~{\footnotesize ~~~~~~~~~~~~~~~~~~~~dh25~{*}~(ajay1~+~gam1{*}x(12))/am1~+~~~~~~~~~~~~~~}{\footnotesize \par}

~{\footnotesize ~~~~~~~~~~~~~~~~~~~~(D1/am1)~{*}~z12~~;}{\footnotesize \par}

~

~{\footnotesize ~~~xh(10)~:=~x(10)~+~(x(13)/am2)~{*}~dh~-~~~~}{\footnotesize \par}

~{\footnotesize ~~~~~~~~~~~~~~~~~~~~~~qom2{*}x(10)~{*}~dh25~-~~~~~~~~~~~~~~~~~~}{\footnotesize \par}

~{\footnotesize ~~~~~~~~~~~~~~~~~~~~~~dh25~{*}~(ajay2~+~gam2{*}x(13))/am2~+~~~~~~~~~~~~~~~~}{\footnotesize \par}

~{\footnotesize ~~~~~~~~~~~~~~~~~~~~~~(D2/am2)~{*}~z22~~;}{\footnotesize \par}

~

~{\footnotesize ~~~xh(11)~:=~x(11)~+~(x(14)/am3)~{*}~dh~-~}{\footnotesize \par}

~{\footnotesize ~~~~~~~~~~~~~~~~~~~~~~qom3{*}x(11)~{*}~dh25~-~~~}{\footnotesize \par}

~{\footnotesize ~~~~~~~~~~~~~~~~~~~~~~dh25~{*}~(ajay3~+~gam3{*}x(14))/am3~+~~~~}{\footnotesize \par}

~{\footnotesize ~~~~~~~~~~~~~~~~~~~~~~(D3/am3)~{*}~z32~~;}{\footnotesize \par}

~

~{\footnotesize ~~~xh(12)~:=~x(12)~+~D1{*}z11~-~~~}{\footnotesize \par}

~{\footnotesize ~~~~~~~~~~~~~~~~~~~~~(~am1{*}qom1{*}x(9)~+~ajay1~+~gam1{*}x(12)~)~{*}~dh~+~~~~~~~~~~~~~~~~~}{\footnotesize \par}

~{\footnotesize ~~~~~~~~~~~~~~~~~~~~~~~qom1{*}(am1{*}gam1{*}x(9)~-~x(12))~{*}~dh25~+~~~}{\footnotesize \par}

~{\footnotesize ~~~~~~~~~~~~~~~~~~~~~~(chi1{*}v{*}x(6)~+~gam1{*}ajay1~+~qgam1{*}x(12))~{*}~dh25~-~}{\footnotesize \par}

~{\footnotesize ~~~~~~~~~~~~~~~~~~~~~~~gam1{*}D1{*}z12;}{\footnotesize \par}

~

~{\footnotesize ~~~xh(13)~:=~x(13)~+~D2{*}z21~-~~~}{\footnotesize \par}

~{\footnotesize ~~~~~~~~~~~~~~~~~~~~(~am2{*}qom2{*}x(10)~+~ajay2~+~gam2{*}x(13)~){*}dh~+~~~~~~~~~~~~~~~~~}{\footnotesize \par}

~{\footnotesize ~~~~~~~~~~~~~~~~~~~~qom2{*}(am2{*}gam2{*}x(10)~-~x(13))~{*}~dh25~+~~~~~~~~~~~~~~~~~}{\footnotesize \par}

~{\footnotesize ~~~~~~~~~~~~~~~~~~~~(-chi2{*}v{*}(x(7)+x(6))~+~gam2{*}ajay2~+~~qgam2{*}x(13))~{*}~dh25~-~~~~~~~~~~~~}{\footnotesize \par}

~{\footnotesize ~~~~~~~~~~~~~~~~~~~~~gam2{*}D2{*}z22;}{\footnotesize \par}

~

~{\footnotesize ~~~xh(14)~:=~x(14)~+~D3{*}z31~-~~~~~~}{\footnotesize \par}

~{\footnotesize ~~~~~~~~~~~~~~~~~~~~(~am3{*}qom3{*}x(11)~+~ajay3~+~gam3{*}x(14)~){*}dh~+~~~~~~~~~~~~~~~~~}{\footnotesize \par}

~{\footnotesize ~~~~~~~~~~~~~~~~~~~~~qom3{*}(am3{*}gam3{*}x(11)~-~x(14))~{*}~dh25~+~~~~~~~~~~~~~~~~}{\footnotesize \par}

~{\footnotesize ~~~~~~~~~~~~~~~~~~~~~(chi3{*}v{*}x(7)~+~gam3{*}ajay3~+~qgam3{*}x(14))~{*}~dh25~-~~~~~~~~~~~~~~~~~}{\footnotesize \par}

~{\footnotesize ~~~~~~~~~~~~~~~~~~~~gam3{*}D3{*}z32;}{\footnotesize \par}

~

~{\footnotesize ~~~~~~~for~ii~in~1..config\_space~~loop~~~~~~}{\footnotesize \par}

~{\footnotesize ~~~~~~~~~x(ii)~:=~xh(ii);~~~~~~~}{\footnotesize \par}

~{\footnotesize ~~~~~~~end~loop;~~}{\footnotesize \par}

~{\footnotesize }{\footnotesize \par}

~{\footnotesize ~end~loop;}{\footnotesize \par}

{\footnotesize end~loop;}{\footnotesize \par}

\end{lyxcode}
~

\bibliographystyle{elsart-num}
\addcontentsline{toc}{section}{\refname}\bibliography{/home/pol/Notes/Bibliography/Physics/physics}

\end{document}